\documentclass[a4paper,amsmath,amssymb, oneside]{article}

\usepackage{mathrsfs}
\usepackage{amsmath}
\usepackage{amssymb}
\usepackage{graphicx}
\usepackage{subfigure}
\usepackage{amssymb}
\usepackage{amsthm}
\pdfoutput=1
\usepackage{pgf}
\usepackage{pdfpages}
\DeclareGraphicsExtensions{.jpeg}

\usepackage{anysize}

\begin{document}

\vskip 1cm
\marginsize{3cm}{3cm}{3cm}{1cm}

\begin{center}
{\bf{\Large The effect of spacers on the performance of Micromegas detectors: a numerical investigation}}\\
~\\

Purba Bhattacharya$^{a*}$, Supratik Mukhopadhyay$^{b}$, Nayana Majumdar$^{b}$, Sudeb Bhattacharya$^{c}$\\
~\\
{\em $^a$ School of Physical Sciences, National Institute of Science Education and Research, Jatni, Bhubaneswar - 752005, India}\\
{\em $^b$ Applied Nuclear Physics Division, Saha Institute of Nuclear Physics, Kolkata - 700064, India}\\
{\em $^c$ Retired Senior Professor, Applied Nuclear Physics Division, Saha Institute of Nuclear Physics, Kolkata - 700064, India}
~\\
~\\
~\\
~\\
~\\
{\bf{\large Abstract}}
\end{center}

Micromegas detector is considered to be a promising candidate for a large variety of high-rate experiments.
Micromegas of various geometries have already been established as appropriate for these experiments for their performances in terms of gas gain uniformity, energy and space point resolution, and their capability to efficiently pave large read-out surfaces with minimum dead zone.  
The present work investigates the effect of spacers on  different detector characteristics of Micromegas detectors having various amplification gaps and mesh hole pitches. 
Numerical simulation has been used as a tool of exploration to evaluate the effect of such dielectric material on detector performance. 
Some of the important and fundamental characteristics such as electron transparency, gain and signal of the Micromegas detector have been estimated.  

\vskip 1.5cm
\begin{flushleft}

\end{flushleft}

\vskip 1.5in
\noindent
{\bf ~$^*$Corresponding Author}: Purba Bhattacharya

E-mail: purba.bhattacharya85@gmail.com

\newpage

\section{Introduction}
\label{sec: introduction}

\begin{figure}[hbt]
\centering
\subfigure[]
{\label{Area-1}\includegraphics[scale=0.3]{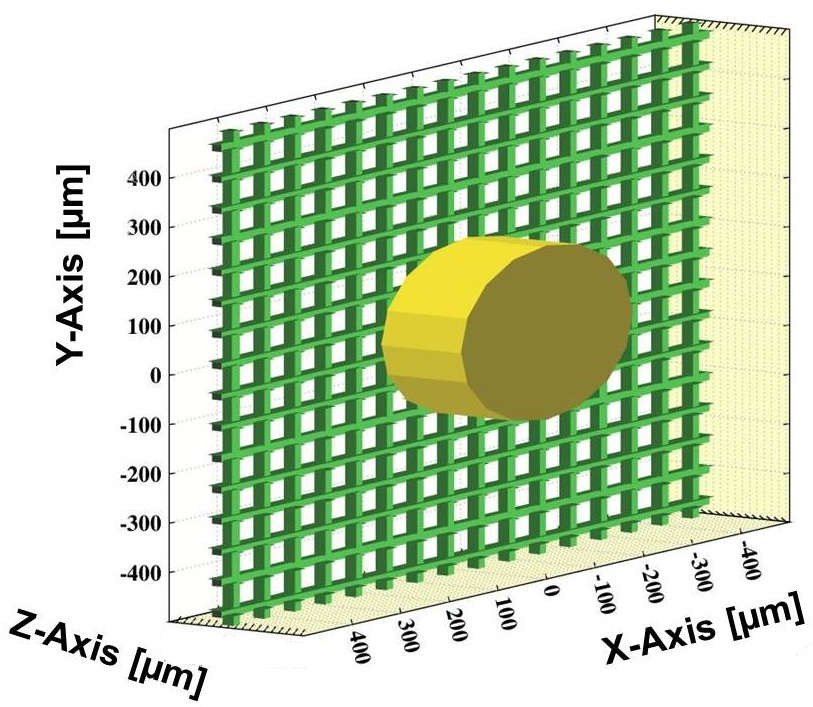}}
\subfigure[]
{\label{Area-2}\includegraphics[scale=0.3]{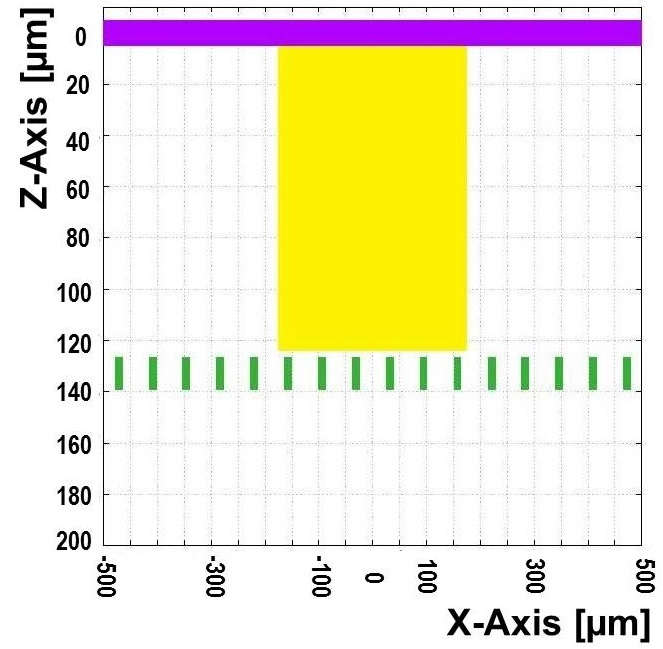}}
\caption{Model of a Micromegas with a cylindrical spacer at the center, (a) 3D view, (b) 2D view.}
\label{Area-Spacer}
\end{figure}

Micro Pattern Gas Detectors (MPGDs) \cite{MPGD}, a recent addition to the gas detector family, have found wide applications for tracking and triggering detectors in different experiments involving astro-particle physics, high energy physics, rare event detection, radiation imaging etc.
For example, GEM and Micromegas detectors are known to offer excellent spatial resolution, high rate capability and single photo-electron time resolution in the several nanosecond range.

The Micromegas (MICRO-MEsh GAseous Structure) \cite{Micromegas}, is a parallel plate device and composed of a very thin metallic micro-mesh, which separates the low-field drift region from the high-field amplification region.
A set of regularly spaced dielectric pillars, called spacer \cite{Spacer}, is required to guarantee the uniformity of the gap between the mesh and the anode plane.
The use of pillars introduces one major drawback.
Particles are not detected at the pillar locations and the sensitivity of the region close to the pillar may be different from the regions where there are no nearby pillars.

The Micromegas \cite{Bulk} detector with an amplification gap of $128~\mu\mathrm{m}$ has been considered to be one of the good choices for a read-out system in different Time Projection Chambers (TPCs) due to its performances in terms of gain uniformity, energy and space point resolution and low ion feedback \cite{TPC2, BULK128, TPC3, TPC4}.
It is also efficient to pave large read-out surfaces with minimum dead zone.
For some experiments involving low pressure operation, Micromegas detectors having larger amplification gaps are more suited \cite{LowPressure1, LowPressure2, LowPressure3}.
Micromegas detectors of a wide range of amplification gaps have been studied for possible application in rare event experiments  \cite{T2Knear, Rare1}. 
Detailed experimental and numerical studies on the basic performance parameters of different Micromegas detector have also been reported on earlier occasions \cite{Purba1, Purba2}.

In this work, numerical simulation has been used as a tool to evaluate the effect of dielectric spacer on the performance of Micromegas detectors. 
The study begins with the extensive computation of the electrostatic field configuration within a given device.
Some of the fundamental properties like gain, electron transmission, signal on the anode plate have been estimated following detailed numerical simulation of the detector dynamics.
The effect of detector geometry, such as amplification gap and mesh hole pitch on the signal has been also studied.

\section{Simulation Tools}
\label{sec: simulation}

The Garfield \cite{Garfield1, Garfield2} simulation framework has been used in the following work.
This framework was augmented in 2009 through the addition of the neBEM (nearly exact Boundary Element Method \cite{neBEM1, neBEM2, neBEM3, neBEM4}) toolkit to carry out 3D electrostatic field simulation.
Besides neBEM, the Garfield framework provides interfaces to HEED \cite{HEED1, HEED2} for primary ionization calculation and Magboltz \cite{Magboltz1, Magboltz2} for computing drift, diffusion, Townsend and attachment coefficients.

\section{Results}
\label{sec: results}

\subsection{Simulation Model}
\label{sec: model}

For numerical simulation, Garfield has been used to model the Micromegas detector.
The simulation model is given in Fig. \ref{Area-Spacer}.
In order to reduce computational complexity, wire elements have been used to model the micro-mesh. 
In addition, in an actual experiment, due to the accumulation of charge, a dielectric spacer is expected to be charged up during operation.
Charging up may lead to gain variations that, depending on the application, can interfere with the correct functioning of the detector.
In our numerical simulation, we have not considered this effect for the present study.
Also the electrostatic deformation of the mesh has not been taken into account in this numerical model.

\begin{figure}
\centering
\subfigure[]
{\label{Field-Without-3D}\includegraphics[scale=0.32]{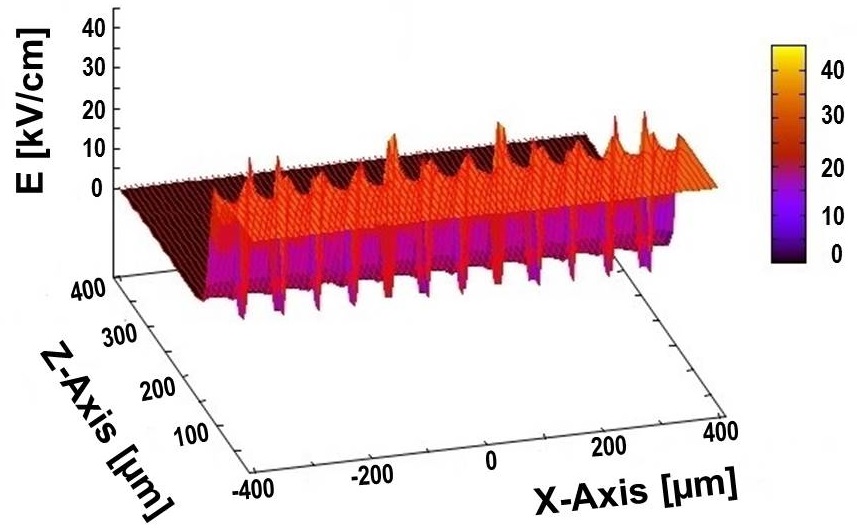}}
\subfigure[]
{\label{Field-With-3D}\includegraphics[scale=0.32]{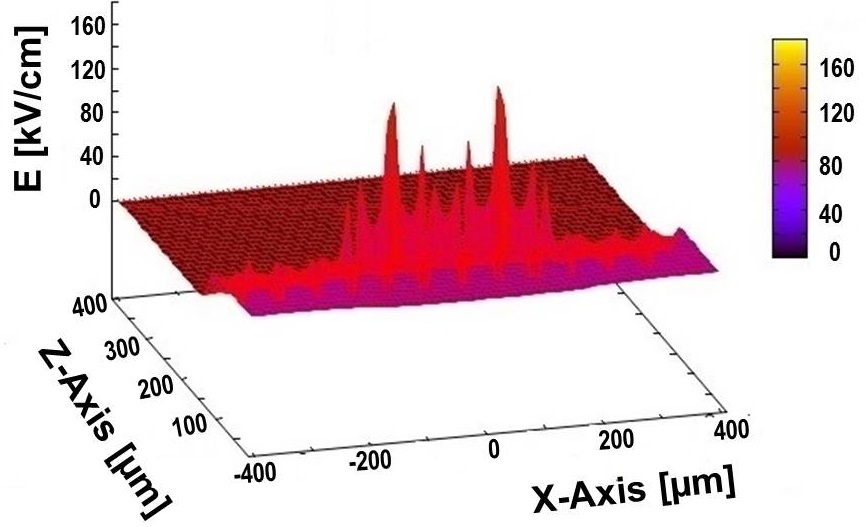}}
\caption{The effect of spacer on electric field in a Micromegas detector having amplification gap of $128~\mu\mathrm{m}$ and pitch of $63~\mu\mathrm{m}$. The field (a) without spacer, (b) with spacer in 3D.}
\label{FieldSpacer}
\end{figure}

The length of the base device along both X and Y directions has been considered to be $2~\mathrm{mm}$.
A cylindrical spacer of diameter $350~\mu\mathrm{m}$ has been placed at the center of the base device.
Thus the pitch between two spacer was maintained at $2~\mathrm{mm}$. 
Then, this base structure has been repeated along both positive and negative X and Y directions to represent a real detector.
The design parameters of the Micromegas detectors considered in this work, are mentioned in Table \ref{Bulkdesign}.

\begin{table}[hbt]
\caption{Parameters of the Micromegas detectors. All detectors have a mesh-wire diameter of $18~\mu\mathrm{m}$.}\label{Bulkdesign}
\begin{center}
\begin{tabular}{|c|c|c|}
\hline
& Amplification gap & Mesh hole pitch \\
& ($\mu\mathrm{m}$) & ($\mu\mathrm{m}$) \\
\hline
MM A & $64$ & $63$ \\
\hline
MM B & $128$ & $63$ \\
\hline
MM C & $128$ & $78$ \\
\hline
MM D & $192$ & $63$ \\
\hline
\end{tabular}
\end{center}
\end{table}

\subsection{Electric Field}
\label{sec: field}

The introduction of a full dielectric cylinder causes significant perturbation resulting in the increase of the field values, particularly in the regions where the cylinder touches the mesh.
Figure \ref{FieldSpacer} shows the effect of such a dielectric spacer on the electric field for the Micromegas with $128~\mu\mathrm{m}$ gap and $63~\mu\mathrm{m}$ pitch. 
As is obvious from the figure, the electric field through the mesh hole near the spacer is also affected by the presence of this dielectric material.
Increase in the electric field values can lead to electric discharges and thus should be avoided, as much as possible, in the detector design.
Therefore, the spacer should occupy the smallest possible volume while keeping the constant amplification gap.

\subsection{Drift Lines, Transmission and Gain}
\label{sec: drift}

\begin{table}
\caption{Electron transmission and gain (without and with spacer) in $\mathrm{Argon}\!-\!\mathrm{Isobutane}$ mixture (${90:10}$). Amplification gap = $128~\mu\mathrm{m}$, pitch = $63~\mu\mathrm{m}$, mesh voltage = 430 V, drift field: 200 V/cm.}\label{Table-Spacer}
\begin{center}
\begin{tabular}{|c|c|c|c|c|}
\hline
 & Track position & \multicolumn{2}{|c|}{Fraction of primary electrons} & Gain\\
\cline{3-4}
 & above mesh ($\mu\mathrm{m}$) & below mesh $\%$ & at anode $\%$ & \\ 
\hline
& $25$ & $87.43$ & $87.43$ & $655$ \\
\cline{2-5}
& $50$ & $86.72$ & $86.72$ & $647$ \\
\cline{2-5}
Without Spacer & $100$ & $87.55$ & $87.55$ & $653$ \\
\cline{2-5}
& $200$ & $86.92$ & $86.92$ & $648$ \\
\cline{2-5}
& $400$ & $87.51$ & $87.51$ & $652$ \\
\hline
\hline
& $25$ & $60.89$ & $56.46$ &  $421$\\
\cline{2-5}
& $50$ & $82.55$ & $80.51$ &  $594$\\
\cline{2-5}
With Spacer & $100$ & $84.25$ & $83.82$ & $620$\\
\cline{2-5}
& $200$ & $84.14$ & $83.79$ & $618$ \\
\cline{2-5}
& $400$ & $84.64$ & $84.36$ &  $624$\\
\hline
\end{tabular}
\end{center}
\end{table}

\begin{figure}
\centering
\subfigure[]
{\label{Drift-Without-25mic}\includegraphics[scale=0.35]{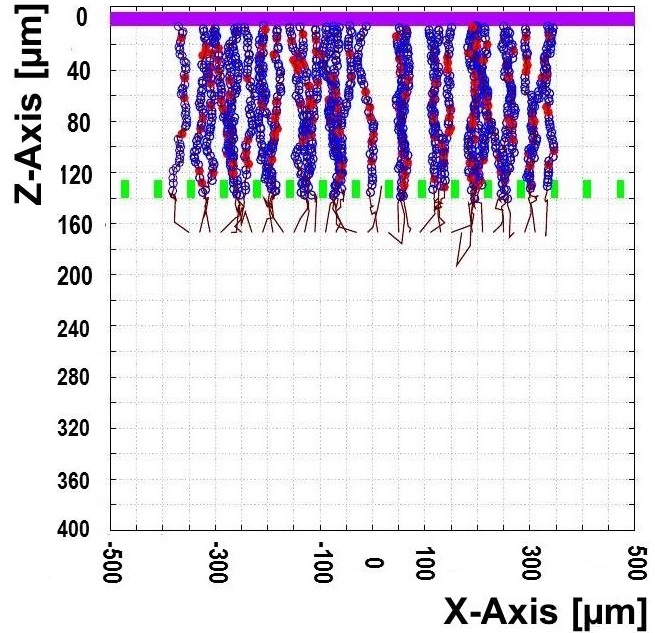}}
\subfigure[]
{\label{Drift-Without-400mic}\includegraphics[scale=0.34]{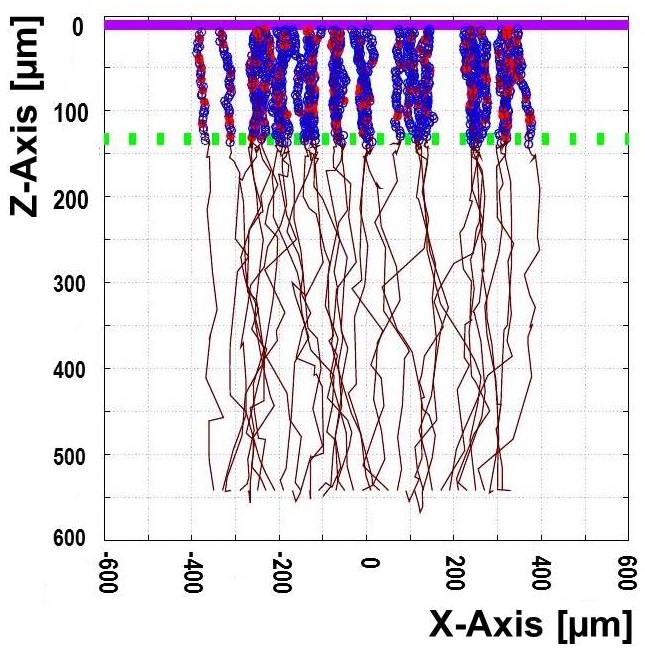}}
\subfigure[]
{\label{Drift-25mic}\includegraphics[scale=0.35]{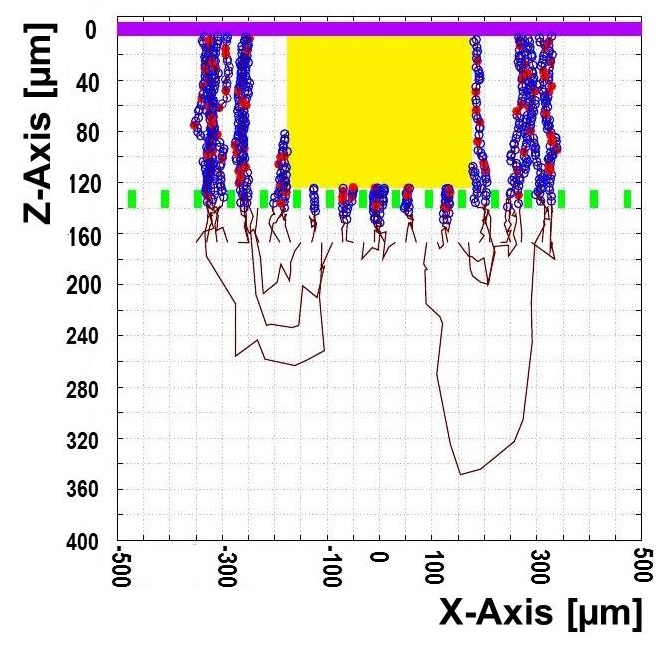}}
\subfigure[]
{\label{Drift-400mic}\includegraphics[scale=0.35]{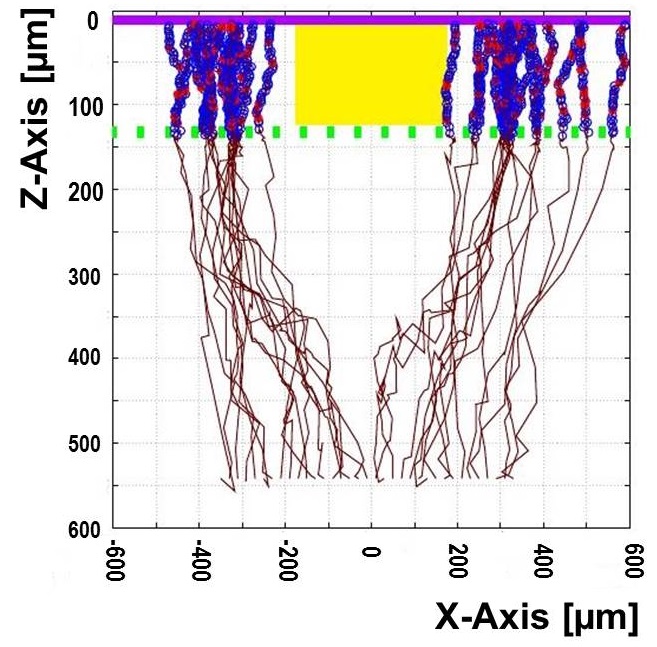}}
\caption{Electron drift lines in $\mathrm{Argon}\!-\!\mathrm{Isobutane}$ mixture (${90:10}$) from a track $25~\mu\mathrm{m}$ above the micro-mesh in case of (a) without spacer and (c) with spacer; from a track $400~\mu\mathrm{m}$ above the micro-mesh in case of (b) without spacer and (d) with spacer. Amplification gap = $128~\mu\mathrm{m}$, pitch = $63~\mu\mathrm{m}$, mesh voltage = $-430~\mathrm{V}$, drift field = $200~\mathrm{V/cm}$. Brown line: electron drift line, blue circle: excitation, red dot: ionization.}
\label{DriftLines-Spacer}
\end{figure}

\begin{figure}
\centering
\subfigure[]
{\label{Time-25mic}\includegraphics[height=0.2\textheight]{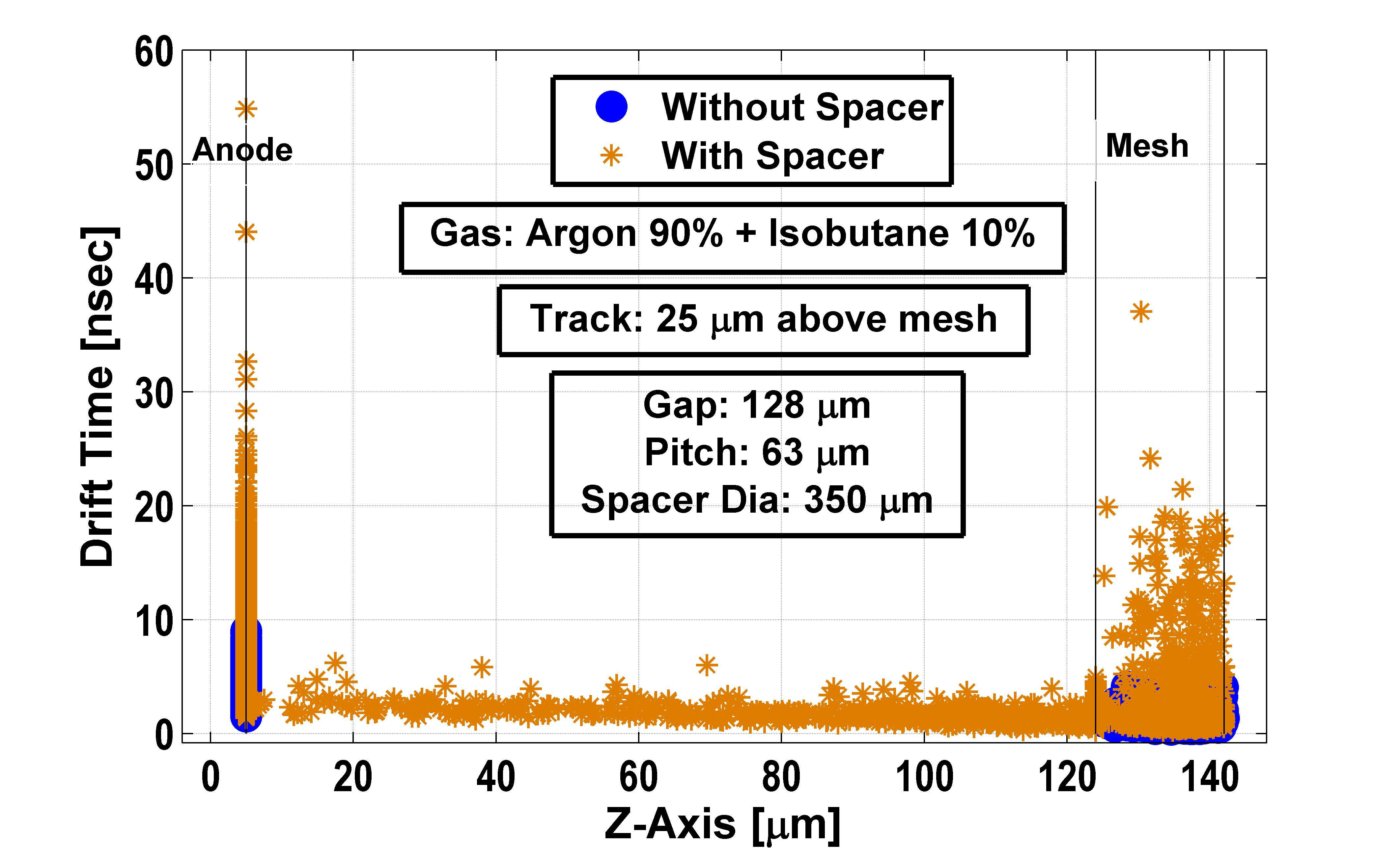}}
\subfigure[]
{\label{Time-400mic}\includegraphics[height=0.2\textheight]{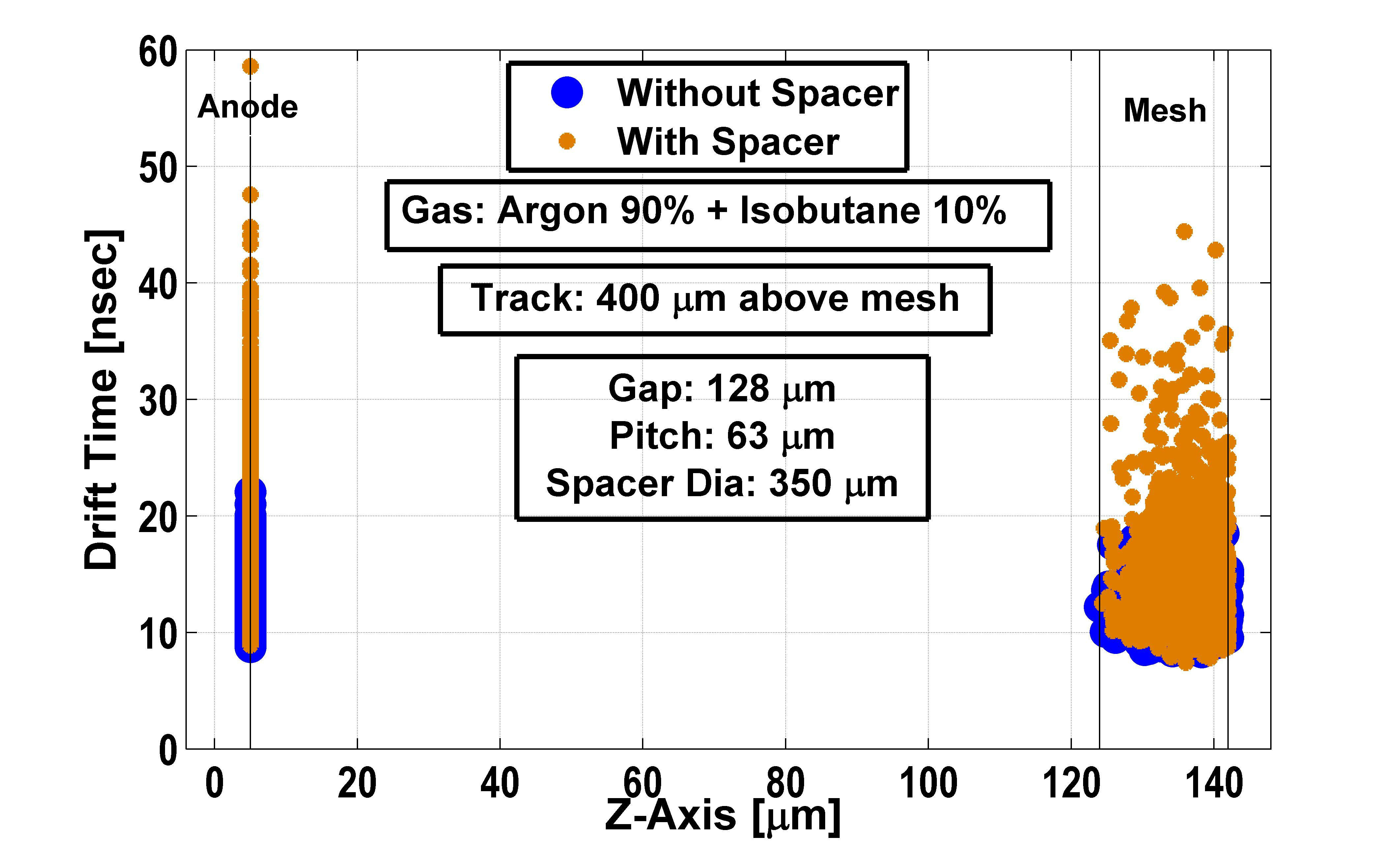}}
\caption{Endpoint (Z-Axis) and drift time of electrons in $\mathrm{Argon}\!-\!\mathrm{Isobutane}$ mixture (${90:10}$) from the track (a) $25~\mu\mathrm{m}$, (b) $400~\mu\mathrm{m}$ above the micro-mesh. Amplification gap = $128~\mu\mathrm{m}$, pitch = $63~\mu\mathrm{m}$, mesh voltage = $-430~\mathrm{V}$, drift field = $200~\mathrm{V/cm}$.}
\label{EndTime}
\end{figure}

To study the effects of the perturbed electric field on different detector characteristics, electron tracks of length $700~\mu\mathrm{m}$ along the X-Axis which extends over the spacer (as shown in Fig. \ref{DriftLines-Spacer}) at different distances from the micro-mesh have been considered.
The electrons from these predetermined tracks are then allowed to drift and multiply in the amplification region.
The plot in Fig. \ref{EndTime} shows the drift time of these electrons versus their end points along Z-Axis.
As seen from Fig. \ref{DriftLines-Spacer}, the drift lines get distorted and a fraction of the electrons in the presence of the spacer take a longer time to reach the anode (Fig. \ref{EndTime}).
From Fig. \ref{DriftLines-Spacer} and \ref{EndTime}, it is also observed that significant number of electrons are lost on the spacer when the track is close to the mesh, resulting in a reduced gain.
From the distant track, at $400~\mu\mathrm{m}$ above the micro-mesh, almost all the electrons do get through, however, circumambulating the spacer and, thus, take longer drift time.
This is the reason why the total gain is not reduced effectively although some space remains insensitive under and close to the spacer (Table \ref{Table-Spacer}).

\begin{figure}[hbt]
\centering
\includegraphics[scale=0.06]{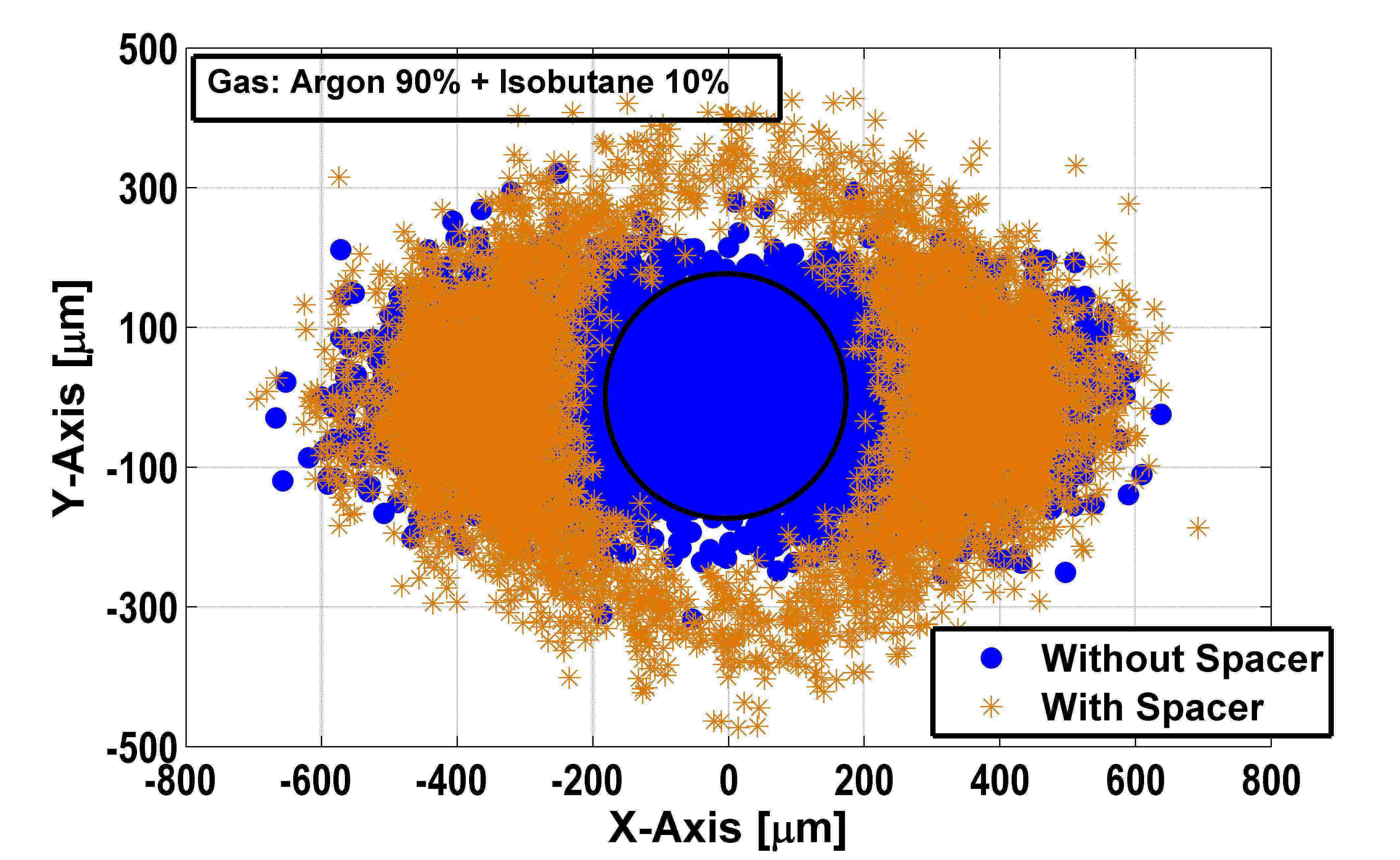}
\caption{Endpoint (X-Axis vs Y-Axis) of electrons in $\mathrm{Argon}\!-\!\mathrm{Isobutane}$ mixture (${90:10}$). Amplification gap = $128~\mu\mathrm{m}$, pitch = $63~\mu\mathrm{m}$, mesh voltage = $-430~\mathrm{V}$, drift field = $200~\mathrm{V/cm}$.}
\label{Endpoint-Spacer}
\end{figure}

\begin{figure}
\centering
\subfigure[]
{\label{ResidualX-Without-25mic}\includegraphics[scale=0.0475]{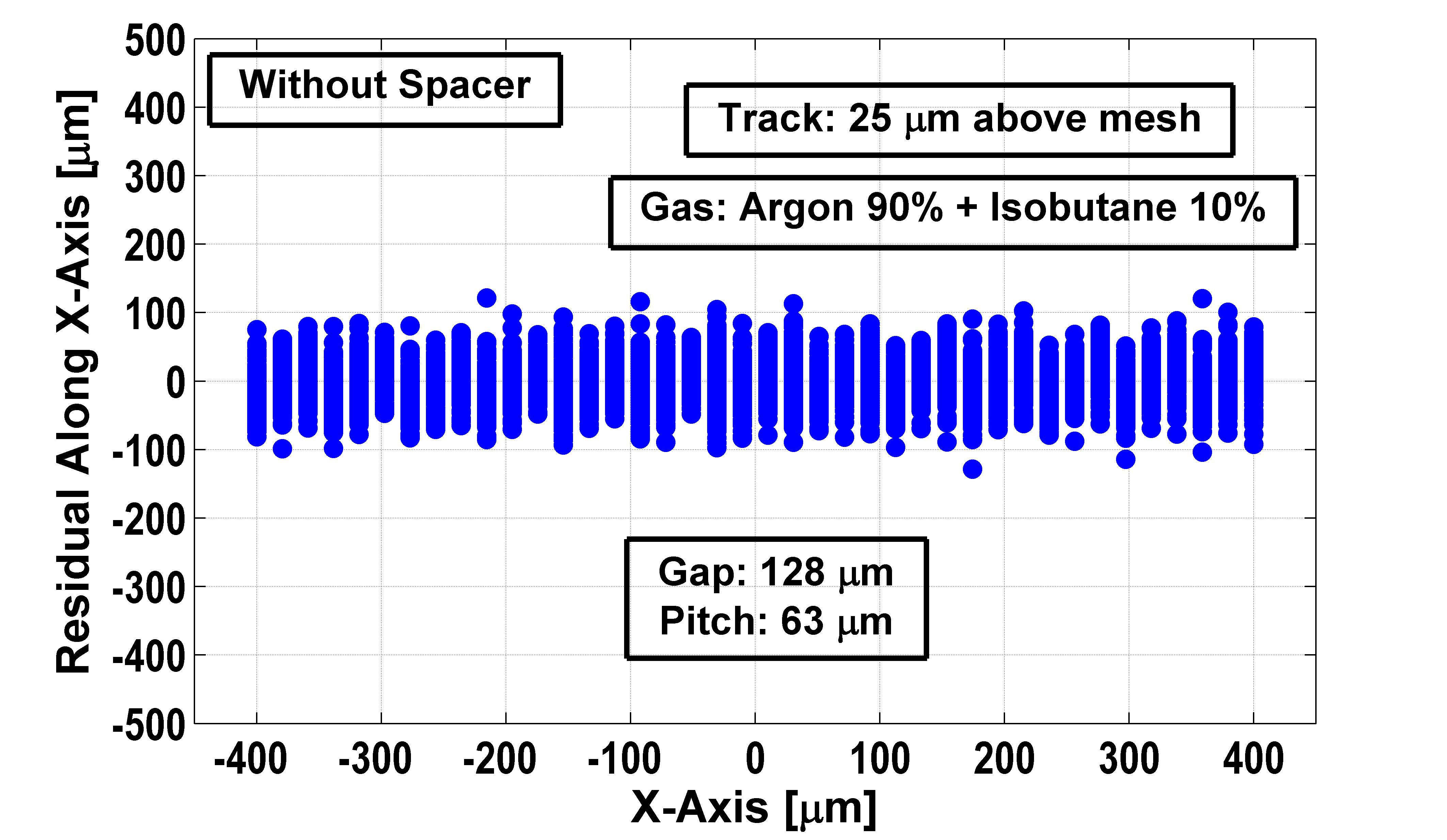}}
\subfigure[]
{\label{Fit-ResidualX-Without-25mic}\includegraphics[scale=0.25]{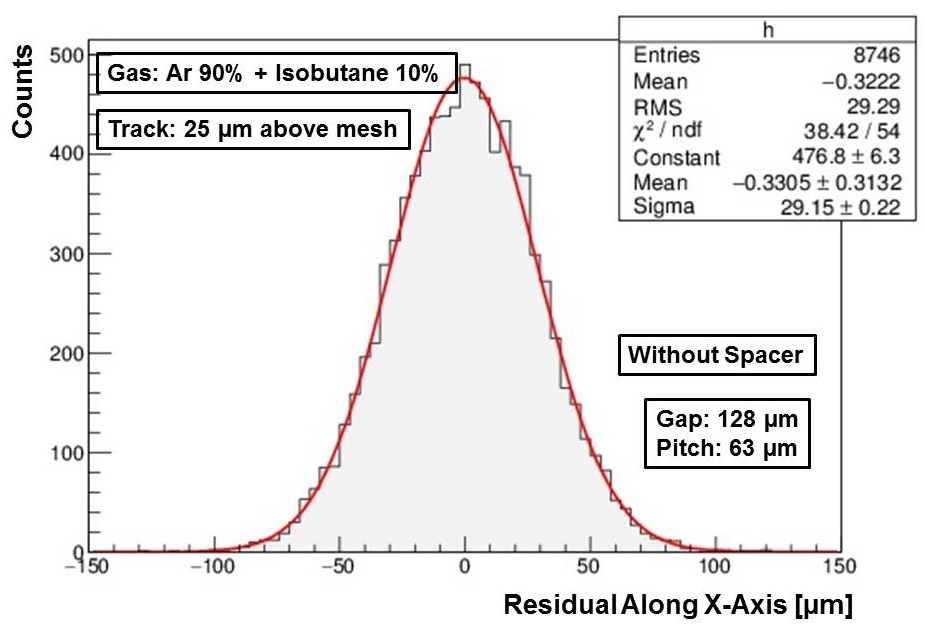}}
\subfigure[]
{\label{ResidualX-Without-400mic}\includegraphics[scale=0.0475]{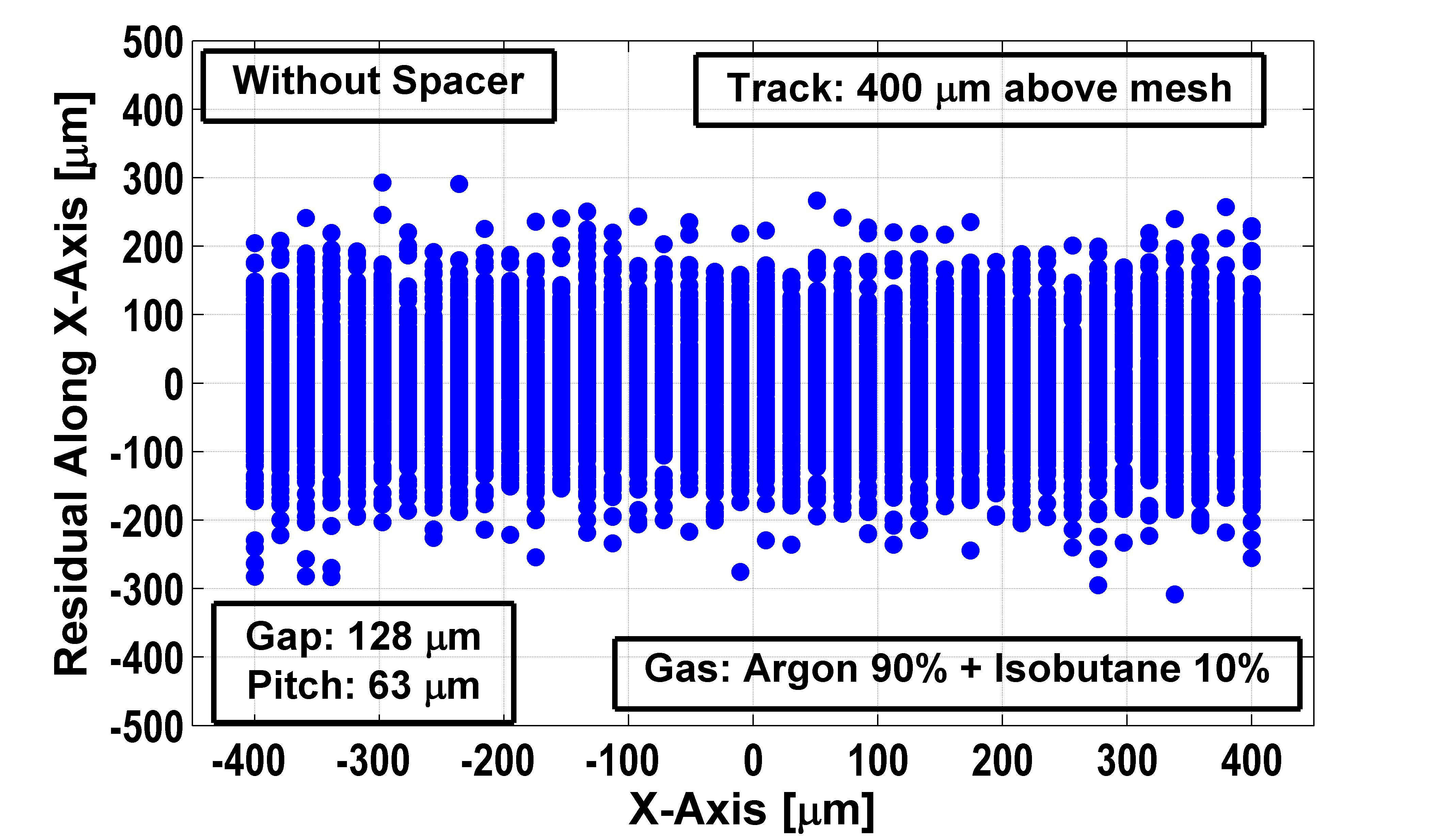}}
\subfigure[]
{\label{Fit-ResidualX-Without-400mic}\includegraphics[scale=0.25]{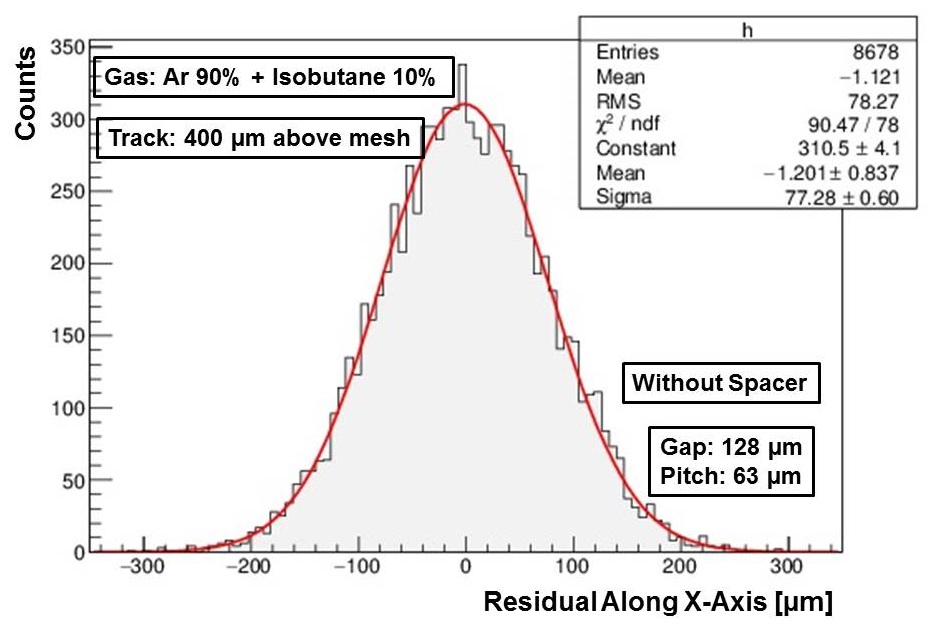}}
\caption{For the Micromegas having amplification gap of $128~\mu\mathrm{m}$ and pitch of $63~\mu\mathrm{m}$, the residual along X-direction without a spacer from a track (a) $25~\mu\mathrm{m}$ and (c) $400~\mu\mathrm{m}$ above the micro-mesh; the residual histogram for the track (b) $25~\mu\mathrm{m}$ and (d) $400~\mu\mathrm{m}$ above the micro-mesh, have been fitted with a Gaussian distribution. Spacer diameter = $350~\mu m$, drift field = $200~\mathrm{V/cm}$.}
\label{Residue-Spacer-1}
\end{figure}

\begin{figure}
\centering
\subfigure[]
{\label{ResidualX-With-25mic}\includegraphics[scale=0.0475]{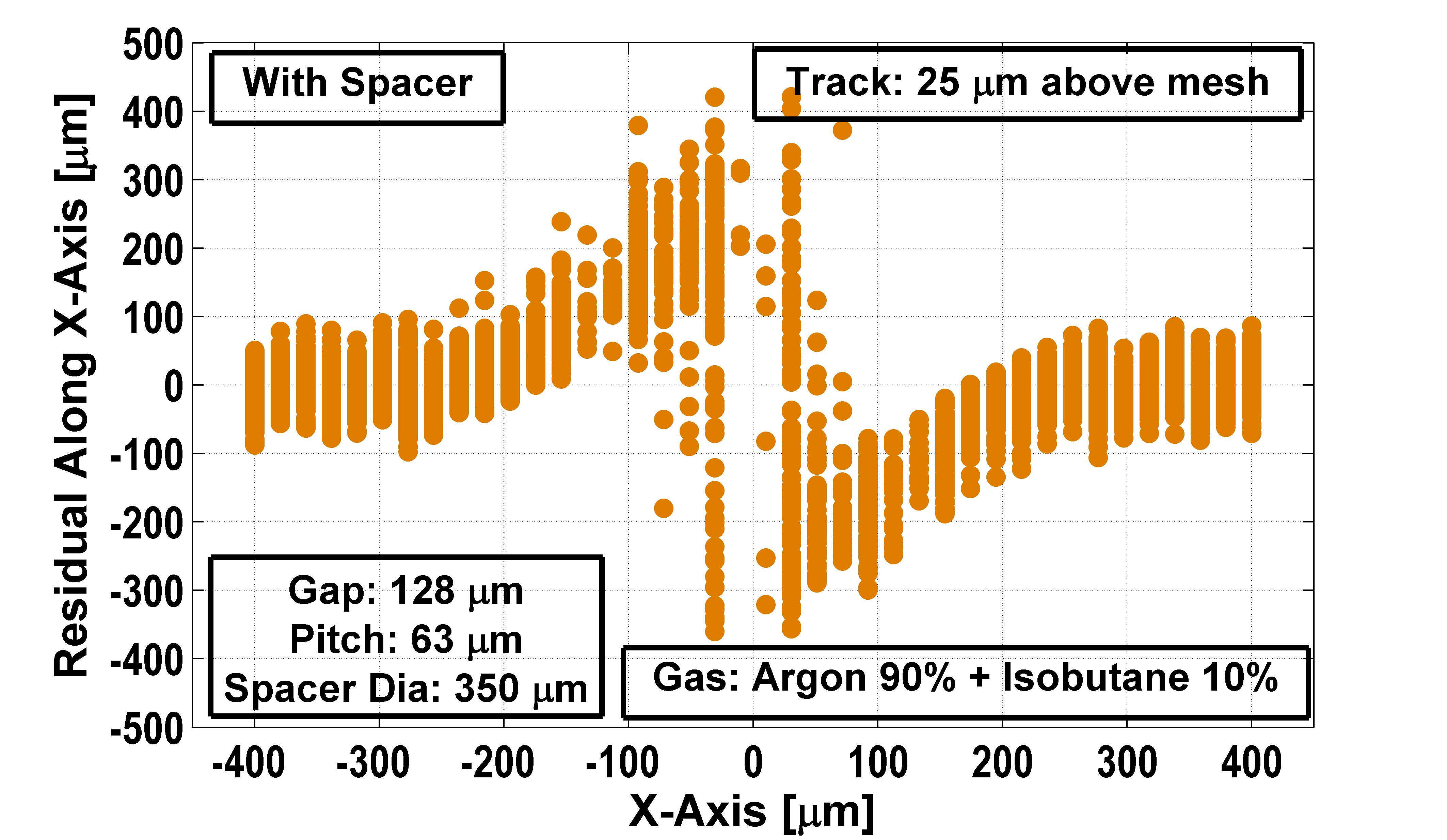}}
\subfigure[]
{\label{ResidualY-With-25mic}\includegraphics[scale=0.0475]{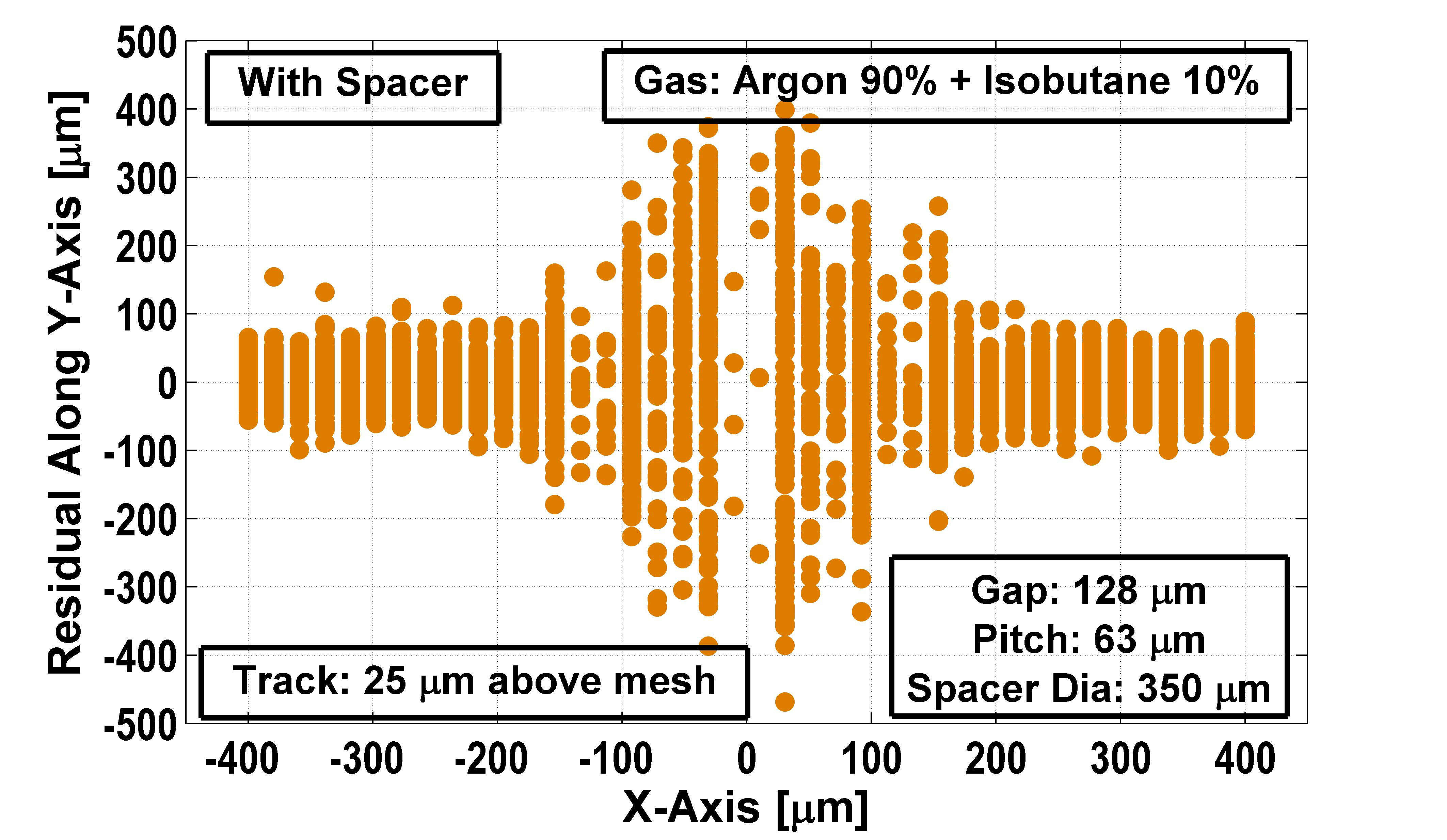}}
\subfigure[]
{\label{ResidualX-With-400mic}\includegraphics[scale=0.0475]{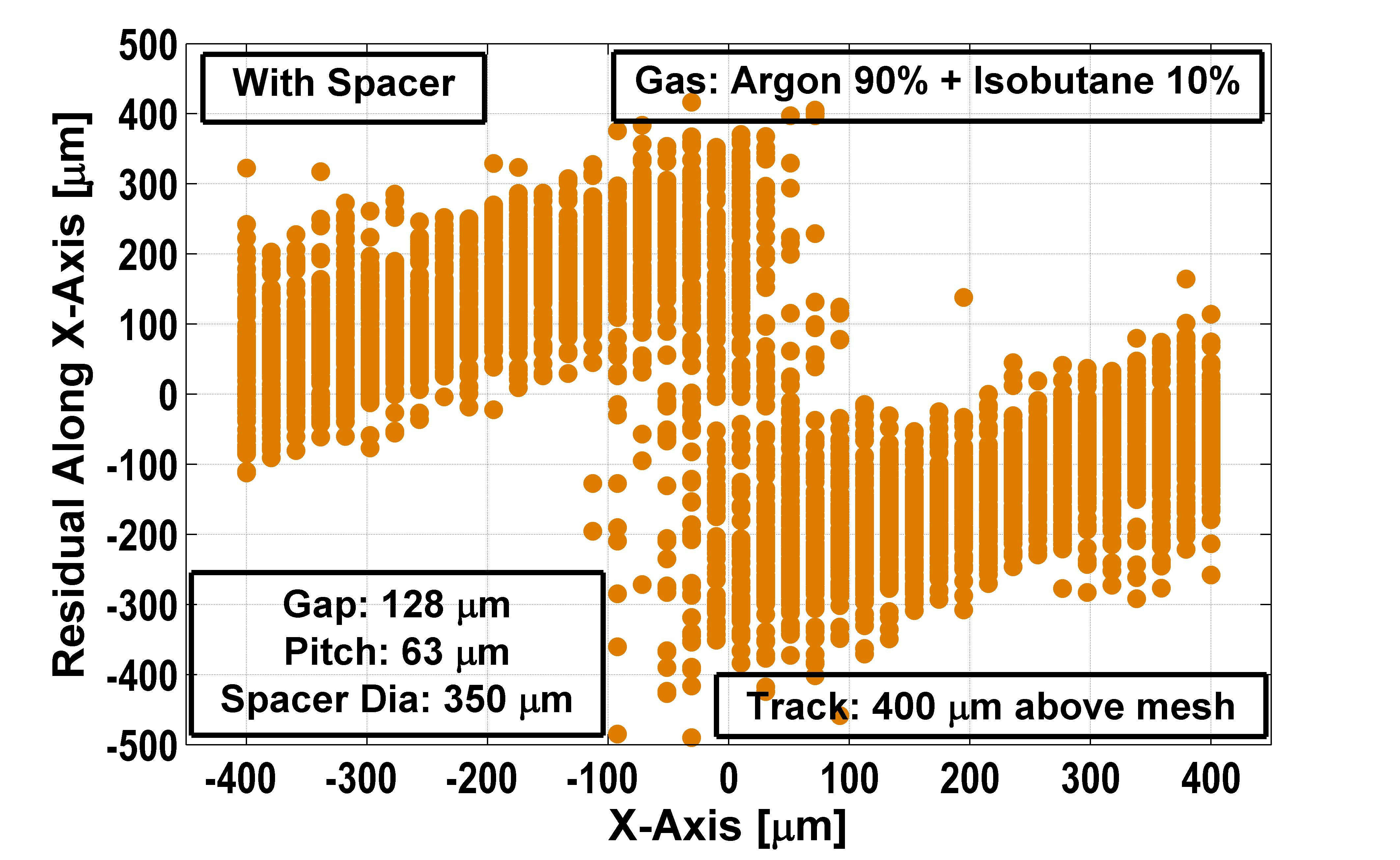}}
\subfigure[]
{\label{ResidualY-With-400mic}\includegraphics[scale=0.0475]{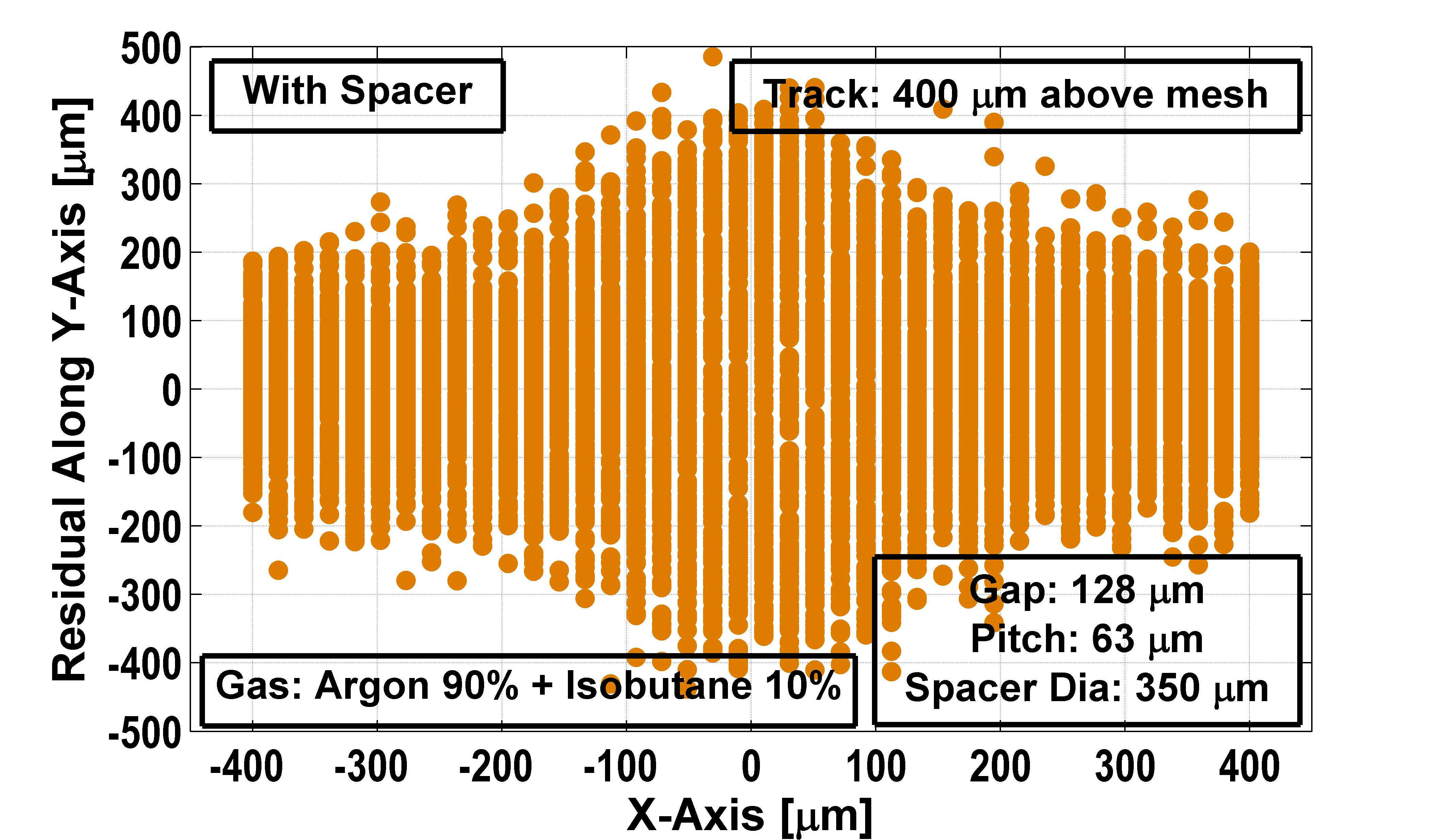}}
\caption{The residual along X-direction with a spacer from a track (a) $25~\mu\mathrm{m}$ and (c) $400~\mu\mathrm{m}$ above the micro-mesh; residual along Y-direction with a spacer from a track (b) $25~\mu\mathrm{m}$ and (d) $400~\mu\mathrm{m}$ above the micro-mesh for the Micromegas having amplification gap of $128~\mu\mathrm{m}$ and pitch of $63~\mu\mathrm{m}$. Spacer diameter = $350~\mu m$, drift field = $200~\mathrm{V/cm}$.}
\label{Residue-Spacer}
\end{figure}

The endpoint of electrons along X and Y-Axes are plotted in Fig. \ref{Endpoint-Spacer}. 
Due to the dead regions introduced by the spacer, the read-out pads below or close to the spacers are found to be affected which leads to inefficiencies in track reconstruction.
In experiments, the residual is calculated to get an idea of the deviation of the measured hit position in the read-out pad with respect to the position of the real track.
Numerically, we have calculated the residual for the pre-defined tracks.
The start and end points of the electrons, which are able to reach the read-out pad, have been noted down and the difference between these two gives an estimation of the ``intrinsic residual''.
We are using the term ``intrinsic residual'' because we have not considered any pad structure (resistive or otherwise) on the anode read-out plane.
It is expected that the real residual will be significantly less than the intrinsic value because of charge sharing and efficient position reconstruction algorithms.
The residual plot for two track positions in absence of spacer are plotted in Fig. \ref{ResidualX-Without-25mic} and Fig. \ref{ResidualX-Without-400mic}, respectively.
The residual histograms for the above mentioned tracks have been fitted using Gaussian distribution as shown in Fig. \ref{Fit-ResidualX-Without-25mic} and Fig. \ref{Fit-ResidualX-Without-400mic}.
As seen from the figures, for the closer track, the rms value of the distribution is $\sim29~\mu\mathrm{m}$ whereas for the track which is $400\mu\mathrm{m}$ above the mesh, the rms value increases to $\sim77~\mu\mathrm{m}$.
This is expected since for the latter case the electrons have to travel a longer path and thus get affected more by diffusion.
In presence of spacer, the deviation around the spacer is large ($\pm450~\mu\mathrm{m}$) as shown in Fig. \ref{ResidualX-With-25mic} and Fig. \ref{ResidualX-With-400mic}, respectively.  
It is quite natural since the diameter of spacer is $350~\mu\mathrm{m}$ and the distorted field above the spacer is such that it deflects the electrons far from its starting position.  
Though in our present study we are considering a track that is at the center of the devices as far as the Y-direction is concerned, the presence of the spacer leads to a deviation also in the Y-direction as shown in Fig. \ref{ResidualY-With-25mic} and Fig. \ref{ResidualY-With-400mic}.
A similar observation has been found experimentally in the test beam run of August 2014 by the ATLAS group working for the MAMMA project.
They have experimentally studied the effect of spacer on single track reconstruction on four identical resistive Micromegas detector with a 2D read-out with $250~\mu\mathrm{m}$ strip pitch in $\mathrm{Ar}$-$\mathrm{CO_2}$ mixture.
For the residual plot they have used three of them to fit the track and plotted the difference of the hit position in the fourth Micromegas with respect to the track position in this Micromegas \cite{Private1}.
It is heartening to note that our simulation results are qualitatively very close to the experimental data.
There are differences in the magnitude which is expected since we are dealing with intrinsic residual while the experiment has different parameter set (detector geometry, resistive anode, gas mixture etc.) and uses well-tested reconstruction algorithms.

\subsection{Signal}

In a medium with perfect conductors and insulators, the current induced by a moving charge ${q}$ onto an electrode can be calculated by means of the Shockley-Ramo theorem.
The current $\mathscr{I}$ that flows into one particular electrode $i$ under the influence of a charge $q$ moving with velocity $v$ can be calculated using the equation 
\begin{eqnarray}
\mathscr{I} = -q{\vec v \times \vec E_w \over V}
\end{eqnarray}
This holds for an arbitrary potential $V$. Here $E_w$, the weighting field, is the field created by raising this electrode to a potential $V$ and grounding all other electrodes, in the absence of charge.

The total charge $Q$ induced on the electrode is given by
\begin{eqnarray}
Q= \int_0^{\Delta t} i(t)dt = q \Delta V_w
\end{eqnarray}
\noindent $\Delta V_w$ is the weighting potential difference across which the charge has drifted.

\begin{figure}
\centering
\subfigure[]
{\label{WeightingField-With1}\includegraphics[scale=0.3]{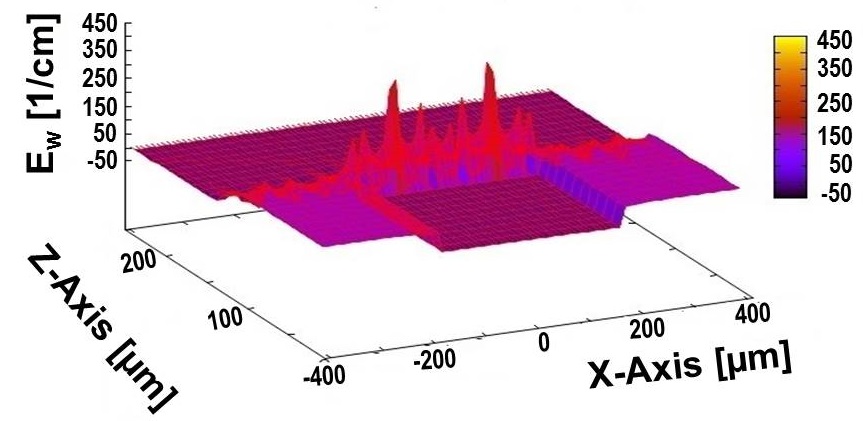}}
\subfigure[]
{\label{WeightingField-Without1}\includegraphics[scale=0.3]{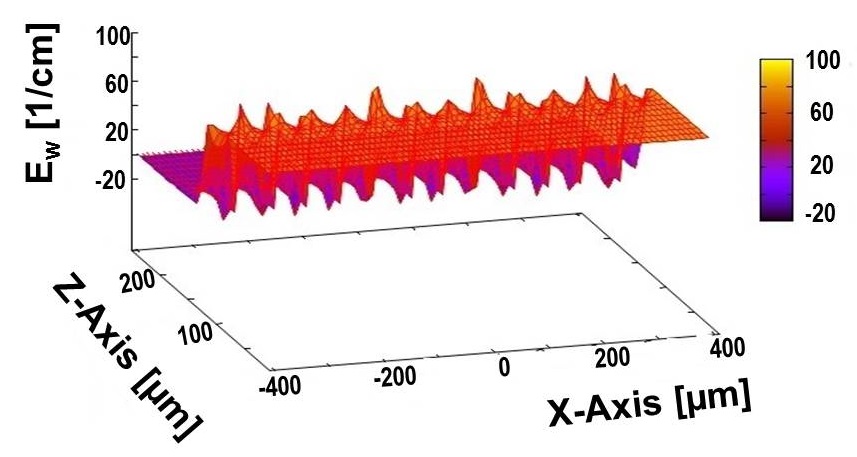}}
\caption{The weighting field (a) in the presence of a spacer, (b) without spacer for a Micromegas detector having an amplification gap of $128~\mu\mathrm{m}$ and a pitch of $63~\mu\mathrm{m}$.}
\label{WeightingFieldSpacer}
\end{figure}

Before calculating the signal, the weighting field and the effect of spacer on it have been estimated.
As in the case of physical electrostatic field, the dielectric spacer also affects the weighting field significantly, as shown in Fig. \ref{WeightingFieldSpacer}.

\begin{figure}[hbt]
\centering
\subfigure[]
{\label{Signal-25mic}\includegraphics[scale=0.0425]{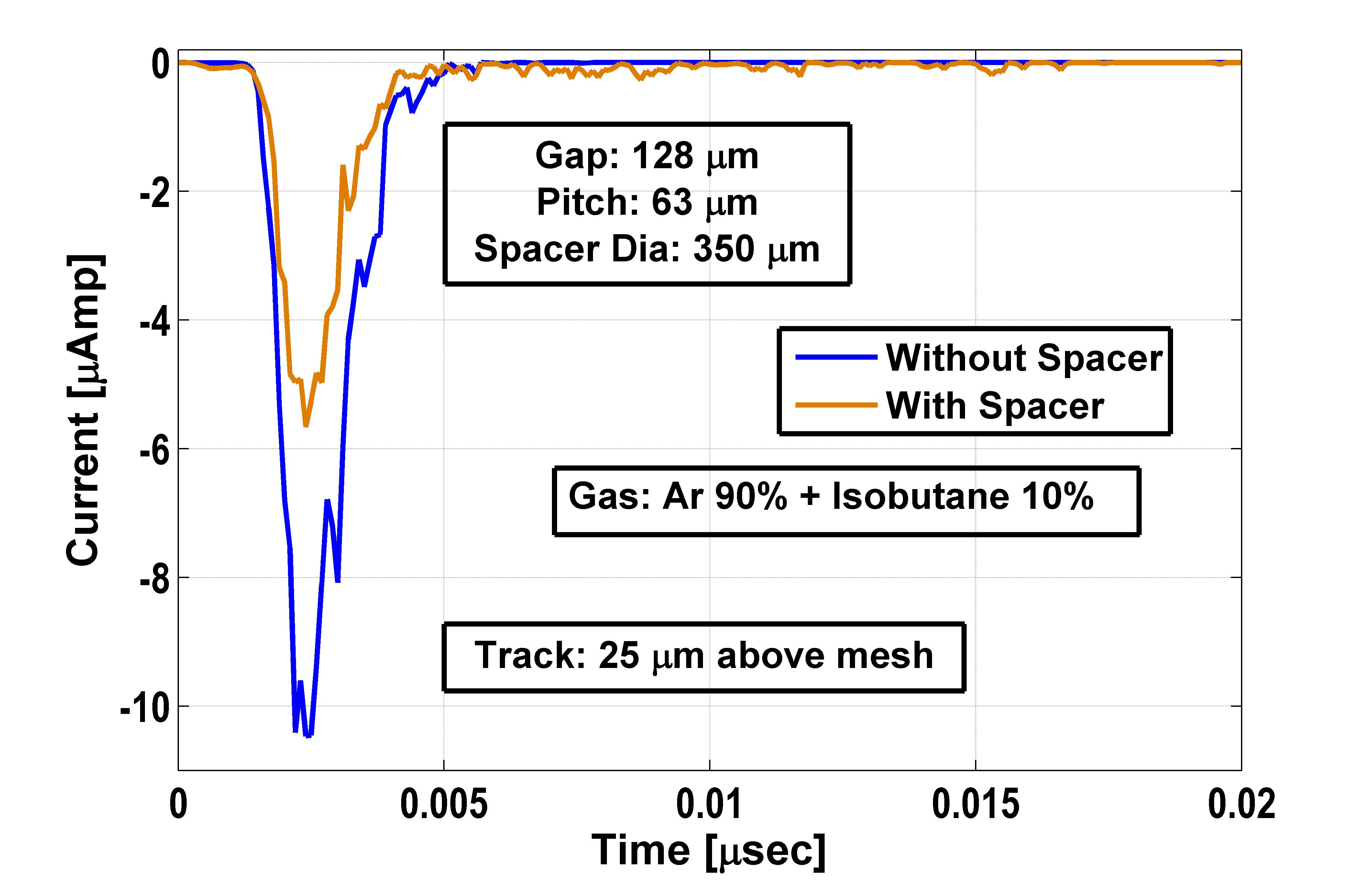}}
\subfigure[]
{\label{Signal-400mic-NoIon}\includegraphics[scale=0.0425]{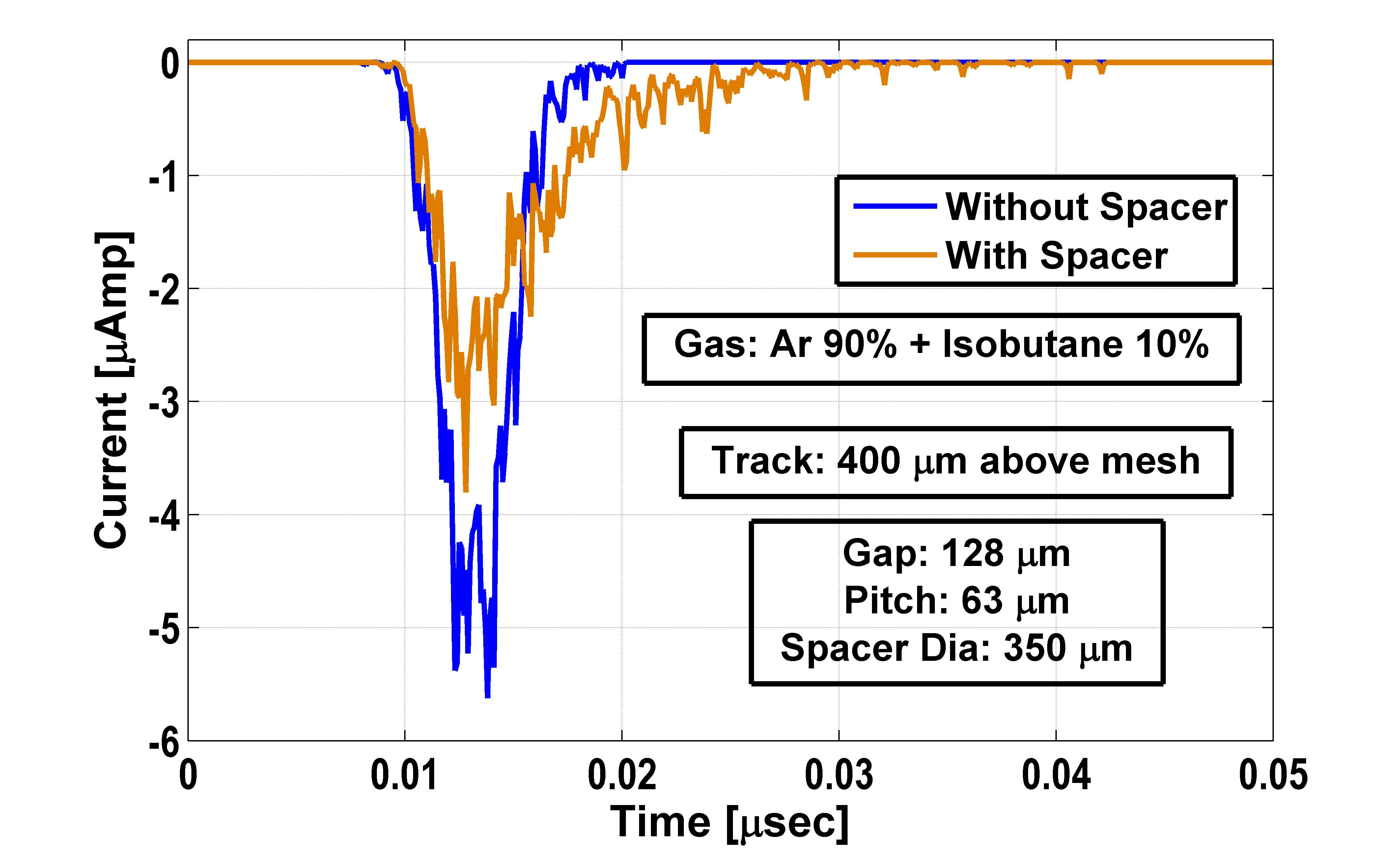}}
\subfigure[]
{\label{Signal-400mic-Ion}\includegraphics[scale=0.0425]{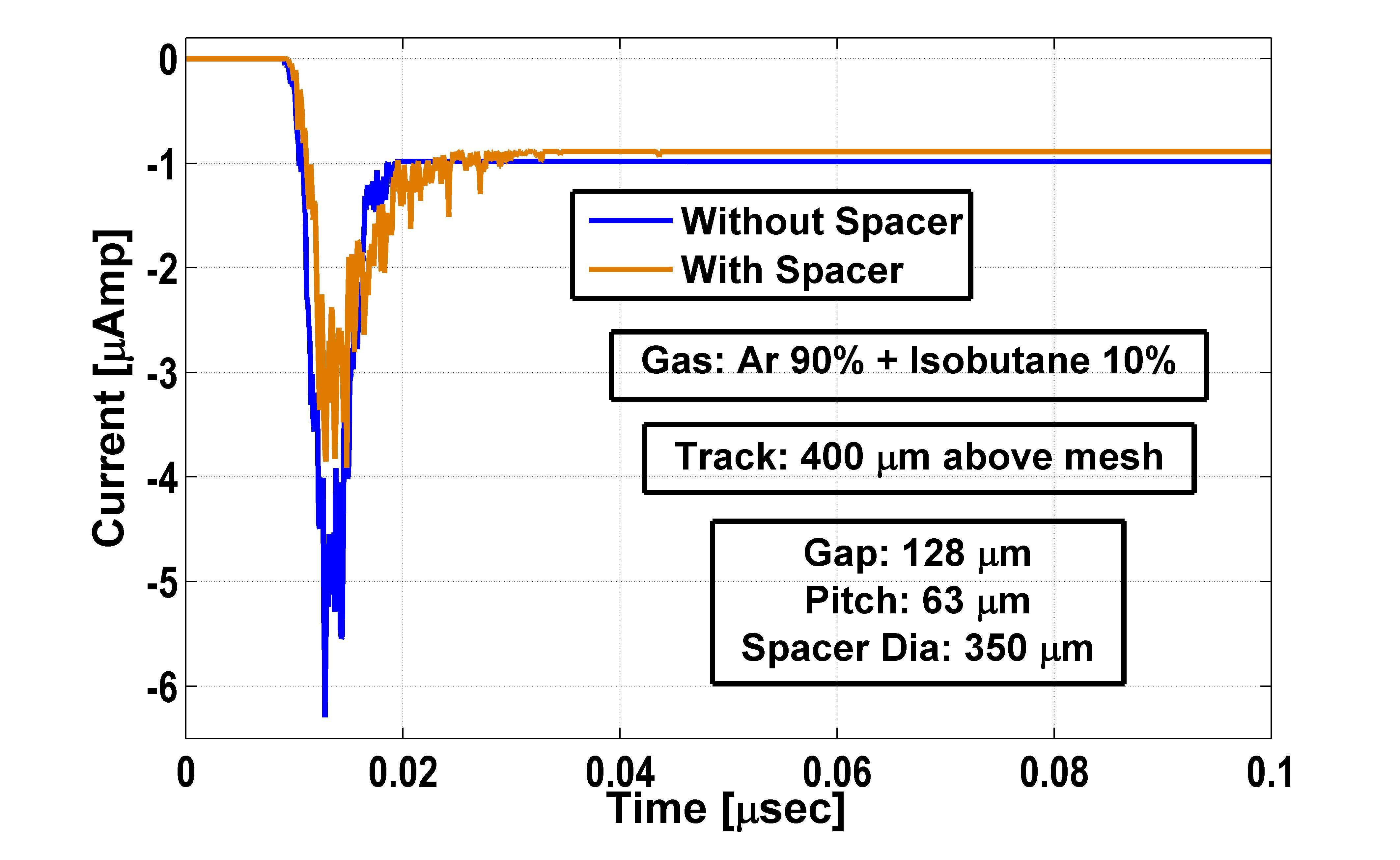}}
\caption{The effect of spacer on cumulative signal for the Micromegas having amplification gap of $128~\mu\mathrm{m}$ and pitch of $63~\mu\mathrm{m}$ in $\mathrm{Argon}\!-\!\mathrm{Isobutane}$ mixture (${90:10}$), due to to all the electrons from a track which is (a) $25~\mu\mathrm{m}$ and (b) $400~\mu\mathrm{m}$ above the micro-mesh.The cumulative signals for both electrons and ions from the tracks $400~\mu\mathrm{m}$ above the micro-mesh are shown in (c). Spacer diameter = $350~\mu\mathrm{m}$, drift field = $200~\mathrm{V/cm}$.}
\label{Signal-Spacer-1}
\end{figure}

Finally, the signal induced on the anode has been simulated.
In this case also, the electron tracks of length $700~\mu\mathrm{m}$ along the X-Axis which extends over the spacer at different distances from the micro-mesh have been considered.
The cumulative electron pulse for 2000 tracks for two track positions are computed by following the avalanche process along the electron drift-line.
For a track nearer to the micro-mesh, due to the reduced gain, the electron signal strength gets affected significantly (Fig. \ref{Signal-25mic}) while for a distant track at $400~\mu\mathrm{m}$ above the micro-mesh, the reduction is much less, in comparison (Fig. \ref{Signal-400mic-NoIon}). 
It is interesting to note that, for both the track positions, the signal profile consists of a long tail resulting from the distorted drift lines. 
The cumulative signals for both electrons and ions from the tracks $400~\mu\mathrm{m}$ above the micro-mesh are shown in Fig. \ref{Signal-400mic-Ion}.
The spikes are due to the electron motion and the long drift time of ions creates the offset current.

\begin{figure}[hbt]
\centering
\subfigure[]
{\label{Signal-Pitch}\includegraphics[scale=0.0425]{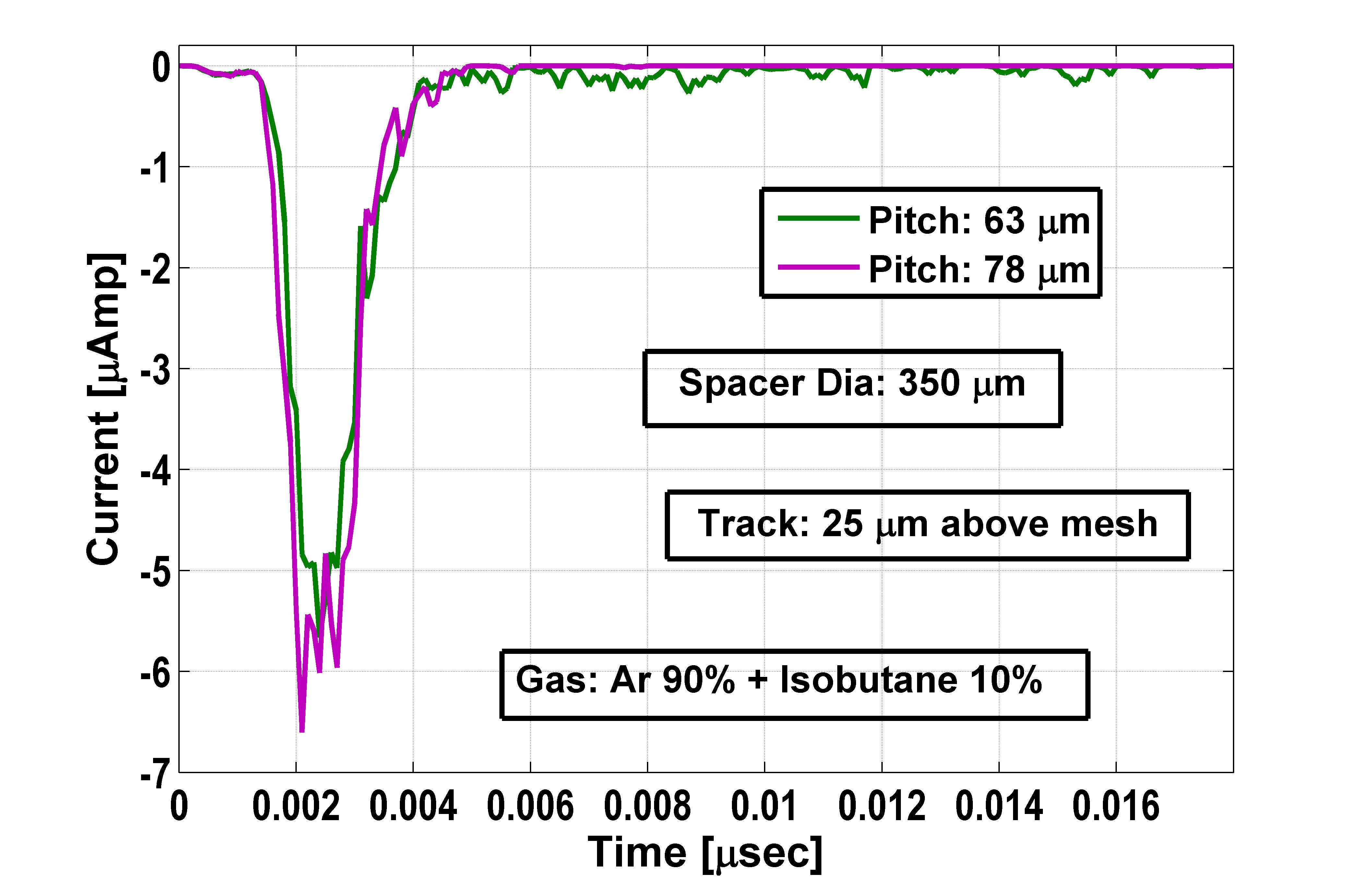}}
\subfigure[]
{\label{Signal-Gap}\includegraphics[scale=0.0425]{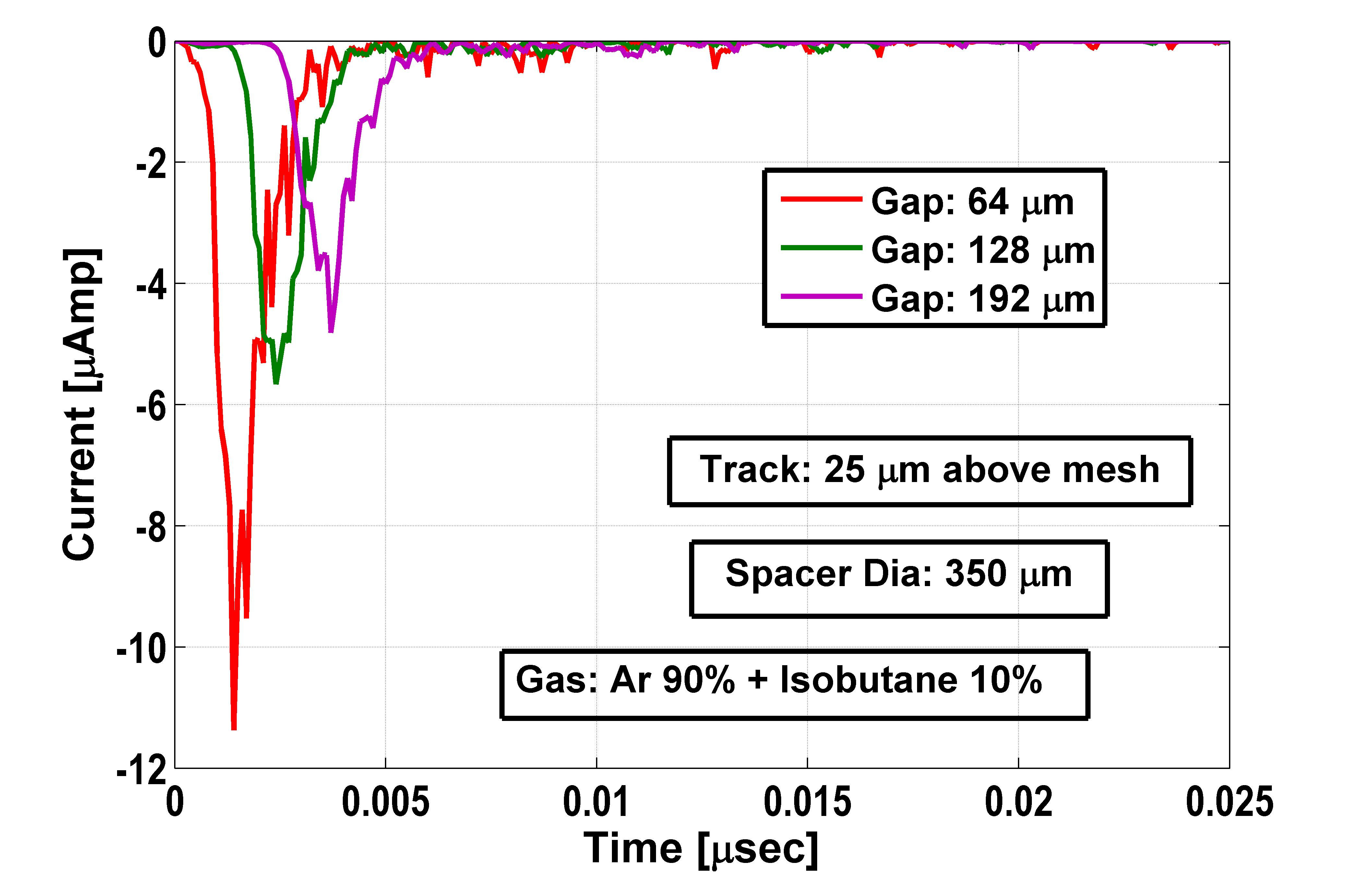}}
\caption{The effect of (a) pitch size and (b) amplification gap on the cumulative signal due to to all the electrons from a track which is $25~\mu\mathrm{m}$ above the micro-mesh. In each of the cases, spacer diameter = $350~\mu\mathrm{m}$, drift field = $200~\mathrm{V/cm}$.}
\label{Signal-Spacer}
\end{figure}

The extent of the electron signal tail, moreover, depends on the mesh hole pitch as shown in Fig. \ref{Signal-Pitch}.
For the smaller pitch, due to the closer proximity among the mesh-wires, the axial field above the mesh, including the spacer region, is perturbed to a greater extent.
As a result, the electric field leading the electrons towards the anode is less for the Micromegas detector having $63~\mu\mathrm{m}$ pitch.
This has two consequences: when the pitch is larger, a) the electrons are more firmly guided towards the anode, b) the electrons face less diffusion.
As a result the signal is found to end earlier for the Micromegas with a larger mesh hole.
In the presence of dielectric spacers, the effect of different amplification gaps on signal has been also studied, as shown in Fig. \ref{Signal-Gap}.
It may be noted here that we have adjusted the simulation parameters such that the gains  are very similar for all the cases under study.
It is observed that for a Micromegas with a smaller gap, the electron signal has an early rise and fall, as expected.
In addition, the magnitude of the electron signal is found to be larger.

\section{Conclusion}

The Garfield+neBEM+Magboltz+Heed combination has been used to simulate the effect of dielectric spacer on the performance of Micromegas detector.
A detailed study of the 3D electric field, electron transmission, gain, endpoint of electrons, drift time, weighting field and signal has been carried out.
It has been observed that the electric field is significantly perturbed due to the introduction of the dielectric material.
The electron transmission and thus the gain is reduced for a track close to the micro-mesh, whereas, the electrons from the distant track are not affected much.
Due to the presence of the spacer, some of the electrons may be lost, or may reach the anode pad at a substantial distance from where it starts.
As a result, the track reconstruction may get significantly affected.
The introduction of dielectric spacer decreases the signal strength, whereas it has a longer tail resulting from the distorted drift lines.
The signal duration and its amplitude, however, also depends on the other parameters of the detector geometry. 

In the presented studies, we have not considered the effects due to charging up of the dielectric material and those due to the accumulation of space charge.
We plan to address these issues in the near future.

\section{Acknowledgment}

This work has partly been performed in the framework of the RD51 Collaboration.
We happily acknowledge the help and suggestions of the members of the RD51 Collaboration.
We sincerely thank Dr. Joerg Wotschack for his useful and relevant comments.  
We would also like to thank Dr. Paul Colas and Dr. David Attie for their valuable suggestions. 
We acknowledge our collaborators from ILC-TPC collaboration for their help and suggestions.
We also thank the reviewers for their valuable comments which have been helpful to improve the manuscript.
Finally, we are thankful to our Institution for providing us with the necessary facilities and IFCPAR/ CEFIPRA (Project No. 4304-1) for partial financial support.

\appendix

\section{Computation of weighting potential and field}

To find this weighting potential, one must solve the Laplace equation for the geometry of the detector, but with artificial boundary conditions. These are: 1) \, the voltage on the electrodes for which the induced charge is to be calculated, is set equal to unity and 2) \, the voltage on all other electrodes are set to zero.
Thus, the weighting potential is not the actual electric potential in the detector, but instead serves as a convenience that allows simple determination of the induced charge on the electrodes of interest by taking differences in the weighting potential at the start and end of the carrier motion.

In neBEM, we have implemented an algorithm that allows the estimation of weighting field at very little computational expense.
In general, for a given device with only five elements, for the physical field we have the following system of algebraic equations:

\begin{equation}
\begin{pmatrix}
K_{11}&K_{12}&K_{13}&K_{14}&K_{15}\\
K_{21}&K_{22}&K_{23}&K_{24}&K_{25}\\
K_{31}&K_{32}&K_{33}&K_{34}&K_{35}\\
K_{41}&K_{42}&K_{43}&K_{44}&K_{45}\\
K_{51}&K_{52}&K_{53}&K_{54}&K_{55}\\
\end{pmatrix}
\begin{pmatrix}
\rho_{c1}\\
\rho_{c2}\\
\rho_{c3}\\
\rho_{c4}\\
\rho_{c5}\\
\end{pmatrix}
=
\begin{pmatrix}
V_1\\
V_2\\
V_3\\
V_4\\
V_5\\
\end{pmatrix}
\end{equation}

\noindent leading to the physical charge distribution as follows:
\begin{equation}
\begin{pmatrix}
\rho_{c1}\\
\rho_{c2}\\
\rho_{c3}\\
\rho_{c4}\\
\rho_{c5}\\
\end{pmatrix}
=
\begin{pmatrix}
I_{11}&I_{12}&I_{13}&I_{14}&I_{15}\\
I_{21}&I_{22}&I_{23}&I_{24}&I_{25}\\
I_{31}&a_{32}&I_{33}&I_{34}&I_{35}\\
I_{41}&a_{42}&I_{43}&I_{44}&I_{45}\\
I_{51}&a_{52}&I_{53}&I_{54}&I_{55}\\
\end{pmatrix}
\begin{pmatrix}
500\\
400\\
0\\
200\\
0\\
\end{pmatrix}
\end{equation}

\noindent where [$\mathbf{I}$] is the inverse of the influence coefficient matrix [$\mathbf{K}$] and the values in the right most column vector represents the boundary conditions in terms of applied voltages.
This column vector may contain zero values to represent either elements on grounded electrodes, or elements belonging to dielectric-dielectric interfaces \cite{neBEM5}.

From the earlier discussions, we know that the weighting field is computed assuming the electrode, for which the signal is being estimated, is raised to unit potential and the rest of the conductors are kept at zero potential.
Thus, for the weighting field within the same device, we have
\begin{equation}
\begin{pmatrix}
\rho_{w1}\\
\rho_{w2}\\
\rho_{w3}\\
\rho_{w4}\\
\rho_{w5}\\
\end{pmatrix}
=
\begin{pmatrix}
I_{11}&I_{12}&I_{13}&I_{14}&I_{15}\\
I_{21}&I_{22}&I_{23}&I_{24}&I_{25}\\
I_{31}&a_{32}&I_{33}&I_{34}&I_{35}\\
I_{41}&a_{42}&I_{43}&I_{44}&I_{45}\\
I_{51}&a_{52}&I_{53}&I_{54}&I_{55}\\
\end{pmatrix}
\begin{pmatrix}
0\\
1\\
0\\
0\\
1\\
\end{pmatrix}
\end{equation}

It may be noted that the influence coefficient matrix [$\mathbf{K}$] remains unchanged for both the cases.
The elements on the conductor on which the induced signal is being investigated have a potential set to 1, whereas for all other elements the value is zero.
As a result, for estimating the weighting field, we can use the same inverted matrix as for the physical field, and the charge density vector corresponding to the weighting field scenario equals to

\begin{equation}
\rho_{w1}~=~I_{11}~*~0~+~I_{12}~*~1~+~I_{13}~*~0~+~I_{14}~*~0~+~I_{15}~*~1
\end{equation}
\begin{equation}
\rho_{w2}~=~I_{21}~*~0~+~I_{22}~*~1~+~I_{23}~*~0~+~I_{24}~*~0~+~I_{25}~*~1
\end{equation}

If we want to study the signal for another set of electrodes, the relevant elements on those electrodes should be 1, while all others should be assigned 0. So, in order to find out the charge densities for a new weighting field estimation, all that is needed is to carry out a matrix multiplication between the already existing inverted matrix and a suitable column vector.


\begin{thebibliography}{00}

\bibitem{MPGD} P. Fonte et al., Plasma Sources Sci. Technol. 19 (2010) 034021

\bibitem{Micromegas} Y. Giomataris et al., Nucl. Instr. Meth. A 376 (1996) 29

\bibitem{Spacer} P. Lazi$\acute{\mathrm{c}}$ et al., Jour. Instr. 6 (2011) P12003

\bibitem{Bulk} I. Giomataris et al., Nucl. Instr. Meth. A 560 (2006) 405

\bibitem{TPC2} S. Anvar et al., Nucl. Instr. Meth. A 602 (2009) 415

\bibitem{BULK128} A. Delbert, Jour. Phy. 308 (2011) 012017

\bibitem{TPC3} A. Sarrat, Nucl. Instr. Meth. A 581 (2007) 175

\bibitem{TPC4} J. Wotschack, Nucl. Instr. Meth. A 640 (2011) 110

\bibitem{LowPressure1} M. Nakhostin, Nucl. Instr. Meth. A 598 (2009) 496

\bibitem{LowPressure2} J. Pancin et al., Jour. Instr. 7 (2012) C03017 

\bibitem{LowPressure3} F. J. Iguaz et al., Jour. Instr. 6 (2011) P07002

\bibitem{T2Knear} N. Abgrall et al., Nucl. Instr. Meth. A 637 (2011) 25

\bibitem{Rare1} L. Ounalli et al., Jour. Phy. 65 (2007) 012017

\bibitem{Purba1} P. Bhattacharya et al., Nucl. Instr. Meth. A 732 (2013) 208

\bibitem{Purba2} P. Bhattacharya et al., Jour. Instr. 9 (2014) C04037

\bibitem{Garfield1} R. Veenhof, online at http://cern.ch/garfield

\bibitem{Garfield2} R. Veenhof, Nucl. Instr. and Meth. A 419 (1998) 726

\bibitem{neBEM1} S. Mukhopadhyay et al., online at http://cern.ch/neBEM

\bibitem{neBEM2} N. Majumdar et al., Nucl. Instr. and Meth. A 566 (2006) 489

\bibitem{neBEM3} S. Mukhopadhyay et al., Eng. Anal. Boundary Elem. 30 (2006) 687

\bibitem{neBEM4} S. Mukhopadhyay et al., Eng. Anal. Boundary Elem. 33 (2009) 105

\bibitem{HEED1} I. Smirnov, online at http://cern.ch/heed

\bibitem{HEED2} I.B. Smirnov, Nucl. Instr. Meth. A 554 (2005) 474

\bibitem{Magboltz1} S. Biagi, online at http://cern.ch/magboltz

\bibitem{Magboltz2} S.F. Biagi, Nucl. Instr. Meth. A 421 (1999) 234

\bibitem{Private1} Private Communication with J. Wotschack

\bibitem{neBEM5} N. Majumdar et al., Jour. Instr. 2 (2007) P09006

\end{thebibliography}
\end{document}